\documentclass[aps,footinbib,twocolumn,showpacs,amsmath,amssymb,superscriptaddress,prb,groupedaddress]{revtex4}
\usepackage{graphicx,amsmath}
\usepackage{epstopdf}
\usepackage{amssymb}
\usepackage{grffile}
\usepackage[usenames]{color}
\usepackage{indentfirst}
\usepackage{float}
\usepackage{color}
\usepackage{mathrsfs}
\usepackage{dcolumn}
\usepackage{bm}
\usepackage[colorlinks=true,bookmarks=false,citecolor=blue,linkcolor=red,urlcolor=blue]{hyperref}
\usepackage{appendix}
\usepackage[T1]{fontenc}
\usepackage{lmodern}

\begin{document}
\title{The geometric impact of the quantum Hall interface on a cone}
\author{Jie Li}
\affiliation{Department of Physics and Chongqing Key Laboratory for Strongly Coupled Physics, Chongqing University, Chongqing 401331, People's Republic of China}
\author{Qi Li}
\affiliation{GBA Branch of Aerospace Information Research Institute, Chinese Academy of Sciences, Guangzhou 510535, China}
\affiliation{Guangdong Provincial Key Laboratory of Terahertz Quantum Electromagnetics, Guangzhou 510700, China}
\author{Zi-Xiang Hu}
\email{zxhu@cqu.edu.cn}
\affiliation{Department of Physics and Chongqing Key Laboratory for Strongly Coupled Physics, Chongqing University, Chongqing 401331, People's Republic of China}
\begin{abstract}
 Recently, quantum Hall interface has become a popular subject of research; distinct from that of the quantum Hall edge, which is constrained by external background confinement, the interface has the freedom to move, likely towards a string-like state. In disk geometry, it was known that the interface energy has an extra correction due to its curvature which depends on the size of the disk. In this work, we analytically calculate the energy of the integer quantum Hall interface on a cone surface which has the advantage that its curvature is more easily adjustable. By tuning the length and curvature of the interface by the cone angle parameter $\beta$, we analyze the dependence of the quantum Hall interface energy on the curvature and verify this geometric correction. Moreover, we find that the tip of the cone geometry has an extra contribution to the energy that reflects on the $u_2,u_4$ term. 

\end{abstract}
\date{\today}
\maketitle

\section{Introduction}
Since the discovery of the quantum Hall (QH) effect, various exotic topologically ordered phases have been uncovered~\cite{Klitzing80,Tsui,Laughlin83}, establishing it as one of the most critical research topics in condensed matter physics. The hallmark experimental signature of the QH effect is its quantized Hall conductance (a transverse response to the electromagnetic field), a characteristic determined by the system's topological properties, and remains unaffected by specific details like impurities and defects, highlighting the topological robustness of the QH state. Beyond the usual descriptions of a QH system's electromagnetic response, the topological state also responds to the geometric manifold in which the electrons reside. For example, the fractional quantum Hall (FQH) state on a torus exhibits topological degeneracy, whereas on a sphere it has a topological shift due to the geometric curvature effect. The geometric responses, including the anomalous viscosity~\cite{Avron, Levay, Read} and the gravitational anomaly~\cite{bradlyn, Can, Abanov}, are lesser known but remain topological features of the QH state. Haldane emphasized that the degree of freedom of the internal geometric leads to the dynamic variation of the guiding-center metric~\cite{Haldane, Haldane1}. Building on earlier pioneering work by Wen and Zee~\cite{Wen}, more efforts have been dedicated~\cite{Can1, Biswas, Ying} to studying the response of FQH states to changes in spatial geometry and topology, such as points of singular curvature in real space or geometries with different topological genus.

In a QH liquid with a boundary, the electrons prefer to stay far away from each other since the Coulomb repulsive interaction, the location of its edge is typically determined by the external confining potential. This potential raises Landau levels near the edge, thereby maintaining the electrons in close proximity to form a droplet. It also influences the edge spectrum, as well as other static and dynamic characteristics of the edge~\cite{wen-book}. The situation becomes significantly different when dealing with interfaces. Take, for example, the magnetic domains within an Ising ferromagnet; here, the domain walls act as interfaces. Because of polarization degeneracy, these domain walls are free to move. Although interfaces in practical scenarios might be fixed by imperfections or other external influences, in an idealized context, there is no analogous potential that confines the interfaces. In this scenario, the interface behaves as an extended string-like object that can move freely in the (2 + 1) dimensional space-time~\cite{McGreevy}. Different QH phases can live in various regions of the sample. A notable example receiving significant attention is the FQH state at $\nu=5/2$ in the GaAs-based quantum well~\cite{Willett1987, Pan1999}, where the two candidate wave functions, the Pfaffian and anti-Pfaffian states are exactly degenerate in the ideal limit (without particle-hole asymmetry and extrinsic effects). In this limit, the pinning effects of the domain walls vanish, rendering them as string-like interfaces. On the one hand, the system's low-energy physics is still governed by area-preserving deformations of the QH liquid~\cite{Cappelli1993-1, Cappelli1993-2}, also known as edge waves~\cite{Stone1991}. In contrast, the string-like nature of the interface significantly alters the excitation spectra, leading to intriguing new physics.  Recently,  Li et al.~\cite{Liqi2021} explored the qualitative differences between edges and interfaces in the disk geometry. Their findings suggest that bosonic interface modes exhibit cubic dispersion, contrasting the linear dispersion observed in edge excitations. However, theoretical forecasts diverge from numerical results at high momentum. Subsequently, the authors of Ref.~\onlinecite{yangku} regard the higher-order correction as the geometric curvature of the interface. In doing so, they derive a more precise low-energy dispersion relation that closely matches the numerical simulations.

Recent experiments have demonstrated that Landau levels of photons can be produced on a cone within a synthetic magnetic field created by continuous photons~\cite{Otterbach,Mittal,Schine16,Schine19, Carusotto,Clark20}, thus establishing the QH system in a new curved space and opening avenues for studying the effects of varying spatial geometry and topology~\cite{Ying, Can1,Liqi2024}, especially the unique curvature at the apex of cone geometry holds significant physical implications. In this case, the boundary separating two distinct states has a more easily tuneable curvature not only by increasing the size of the system but also by changing the angle of the cone geometry.~\cite{wan16,Yangku17,Hughes19,zhu20,Nielsen20,Nielsen21,Gefen21,Estienne,Regnault,Liqi2021} Based on earlier analytical studies on QH systems in disk geometry~\cite{Ciftja03,Ciftja09,Ciftja11,Bentalha2014}, we investigate the geometric impact of the QH interface on a cone using analytical calculations and numerical integration. In the following of this paper, Sec.~\ref{sec:Analy} is derived analytical formulas for the Coulomb interaction and hard-core interaction within the IQH system in cone geometry. Sec.~\ref{sec:Numerical} compared the analytical results with those obtained from exact diagonalization. When the interactions are attractive, the boundary is considered as the interface between the quantum Hall liquid and the vacuum. Due to the non-uniform density at the apex of the cone, there is an additional energy contribution at tip, which we calculated using the Hartree-Fock approximation. After considering the tip energy, we analyzed the dependence of the QH interface energy on curvature. Finally, we summarize our work in Sec.~\ref{sec:Conclusion}.

\section{Analytical results of the energy}
\label{sec:Analy}
\begin{figure}
  \includegraphics[width=8.5cm]{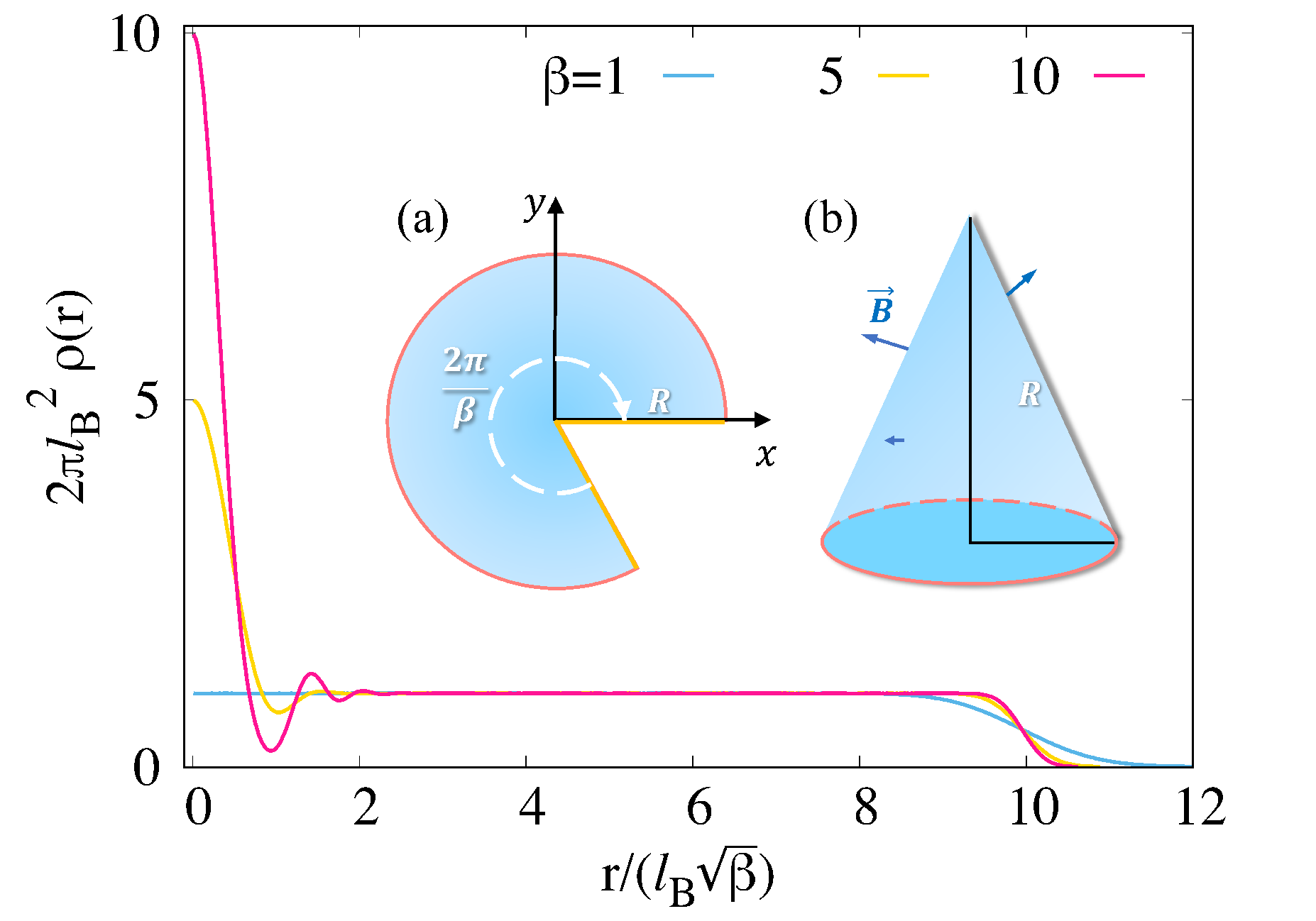}
  \caption{\label{fig:cone}The radial density for 50 electrons on a cone. The two inserted plots are the sketch of a cone composed of planar disks. (a) Cut a $1/\beta$ sector from a disk with a radius of $r$, and stick the two edges (yellow solid line) together to form a cone. (b) For a system of $N$ particles, $r=\sqrt{2N\beta}l_B$, $R=\sqrt{2N/\beta}l_B$. The bottom of the cone forms a new circle with a radius of $R$, this brings a new curvature $k=\sqrt{\beta/(2N)}/l_B$. }
\end{figure}

As sketched in Fig.~\ref{fig:cone}(a)(b), the cone can be made up of a disk. A sector is cut from the disk at an angle of $2\pi / \beta$, and the two edges obtained are glued together. The bottom of the cone forms a new circle based on the arc length of the sector. Therefore, the parameter $\beta$ can control the length and curvature of the boundary with a fixed surface area at a given number of electrons and the filling factor. When $\beta=1$, it returns to disk geometry; as $\beta\to \infty$, it becomes a thin cylinder. Assuming a uniform magnetic field passes through the surface, and the effective single-particle Hamiltonian is $H = \frac{1}{2m}(\mathbf{p} -e\mathbf{A}/c)^2$. In cone geometry, the position of the electron is defined by complex coordinates $\mathbf{z}=(x+iy)/l_B =(r/l_B)e^{i\theta}$, where the angle $\theta$ is confined to the interval $[0,2\pi/\beta]$ and $l_B=\sqrt{\hbar c/eB}$ is the magnetic length. In the lowest Landau level with a symmetric gauge, the solution of single particle wave function is:
\begin{eqnarray}
    \psi_{\beta m}(\vec{r}) = \sqrt{\frac{{\beta}}{2\pi l_B^2 2^{\beta m} (\beta m)!}}e^{i\beta m\theta} (r/l_B)^{\beta m} e^{-r^2/4l_B^2}.
\end{eqnarray}
In this case, the wave function of a $\nu = 1$ IQH state is~\cite{Ying}
\begin{align}
     \Psi (\left \{ z \right \}  )=\prod_{i<j}^{} (z_{i}^{\beta}-z_{j}^{\beta})\exp(-\sum_{k}^{}\frac{\left | z_{k}  \right | ^{2}}{4l_B^2}),
\end{align}
in disk geometry with $\beta = 1$, this state has a uniform density $\rho_0 = 1/2\pi l_B^2$ in the bulk. While $\beta > 1$, as shown in Fig.~\ref{fig:cone}, there are strong fluctuations within a certain range in a small area near the tip of the cone. As $\beta$ increases, the accumulation of charge at the tip leads to the formation of an interface between the charge density wave (CDW) near the tip and the QH state~\cite{Liqi2024}. In the case of integer filling, the density remains unaffected by the details of the electron-electron interaction because each orbit is fully occupied by a single electron.

Following previous studies, to create an interface rather than a QH edge, we use the attractive forces between the electrons, eliminating the necessity of a confinement potential. For a $N$-electron system, we examine the Coulomb interaction represented as $-\frac{e^2}{\epsilon|\vec{r_1}-\vec{r_2}|}$ where $\epsilon$ is permittivity and also the model interaction with a negative $V_1$ Haldane pseudopotential. For the Coulomb interaction, this interaction is separable into two parts, namely, the direct interaction energy, denoted as $E_{dir}$, and the exchange interaction energy $E_{ex}$. We aim to compute both $E_{dir}$ and $E_{ex}$ analytically in the following.
 
\textit{Direct interaction}. For negative Coulomb interaction, the direct interaction energy $E_{dir}$ is defined as
\begin{equation}\label{vee_dir}
    E_{dir}  = -\frac{e^2}{2\epsilon} \int d^2r_1 \rho(\vec{r_1})  \int d^2r_2 \rho(\vec{r_2}) \frac{1}{|\vec{r_1} - \vec{r_2}|}.
\end{equation}

For $N$ electrons at filling $\nu = 1$, the attractive interaction forces the electrons to form a cluster which does not differ from the IQH statem, namely the electrons occupy consecutive $N$ orbits. Therefore, the electron density at position $\vec{r}$ is $\rho(\vec{r}) = \sum_{m=0}^{N-1} |\psi_{\beta m}(\vec{r})|^2$. After applying a Fourier transform to the $\frac{1}{|\vec{r}_1 - \vec{r}_2|}$ and then performing the integration over the angular variables, Eq.~\eqref{vee_dir} could be simplified in to a one-dimensional integration form
\begin{eqnarray}
    E_{dir} = -{\frac{e^2}{\epsilon l_B} \frac{1}{2} \int dq e^{-q^2} \left[\sum_{m=0}^{N-1}L_{(\beta m)}(q^2/2)\right]^2 }.
\end{eqnarray}
where $L_{m}(x)$ are the Laguerrel polynomials. This could be easily calculated numerically for any system size.

\textit{Exchange interaction}. In a same manner, the exchange interaction energy $E_{ex}$ is
\begin{equation}
    E_{ex}  = \frac{e^2}{2 \epsilon} \int d^2r_1   \int d^2r_2 |\rho(\vec{r_1}, \vec{r_2})|^2 \frac{1}{|\vec{r}_1 - \vec{r}_2|}.
\end{equation}
Here we need to use the square of the two-particle density function
\begin{align}
   & \nonumber |\rho(\vec{r_1}, \vec{r_2})|^2  = |\sum_{m=0}^{N-1} \psi_{\beta m}^*(\vec{r_1}) \psi_{\beta m}(\vec{r_2})|^2.
\end{align}
The computation of $E_{ex}$ can be split into two components, denoted as $E_{ex}^{1}$ and $E_{ex}^{2}$. Particularly, $E_{ex}^{1}$ includes the terms that share an identical summation index,
\begin{align}
    \nonumber E_{ex}^{1} = &\frac{\beta^2 e^2 \rho_0^2}{2\epsilon} \int d^2r_1 \int d^2r_2   \frac{e^{-(r_1^2+r_2^2)/2l_B^2}}{|\vec{r}_1 - \vec{r}_2|} \\
    &\sum_{m=0}^{N-1} \left[\frac{1}{(\beta m)!} (\frac{r_1r_2}{2l_B^2})^{(\beta m)}\right]^2
\end{align}
and $E_{ex}^{2}$ includes the terms that share different summation index $m \ne m'$,
\begin{align}
    \nonumber E_{ex}^{2} =& \frac{\beta^2 e^2 \rho_0^2}{2\epsilon} \int d^2r_1 \int d^2r_2   \frac{e^{-(r_1^2+r_2^2)/2l_B^2}}{|\vec{r}_1 - \vec{r}_2|} \\
    &\sum_{m \neq m'}^{N-1} \sum_{m'= 0}^{N-1} \frac{e^{i\beta(m-m')(\theta_2-\theta_1)}}{(\beta m)!(\beta m')!} (\frac{r_1 r_2}{2l_B^2})^{\beta (m+m')}.
\end{align}
Similarly, by utilizing the Fourier transform of $\frac{1}{|\vec{r}_1 - \vec{r}_2|}$ and executing the angular integration, each of these terms can be described as a one-dimensional integral
\begin{eqnarray}
    E_{ex}^{1} = {\frac{e^2}{2 \epsilon l_B } \int dq e^{-q^2} \sum_{m=0}^{N-1} \left[L_{\beta m}(q^2/2)\right]^2 }
\end{eqnarray}
and
\begin{align}
   \nonumber E_{ex}^{2} = & \frac{e^2}{\epsilon l_B } \sum_{m = m'+1}^{N-1} \sum_{m'=0}^{N-2} \frac{(\beta m')!2^{\beta m'}}{(\beta m)!2^{\beta m}} \\
   & \int dq q^{2\beta (m-m')} e^{-q^2}\left[ L_{\beta m'}^{\beta(m-m')} (q^2/2) \right]^2.
\end{align}

To sum up, the complete energy arising from the Coulomb interaction among electrons is the aggregate of the aforementioned direct and exchange interaction energies:
\begin{eqnarray}\label{eq:eevc}
    E_{V_C} =  E_{dir} + E_{ex}^{1} + E_{ex}^{2}.
\end{eqnarray}

According to Ref.~\onlinecite{Liqi2021}, we employ the Haldane $V_1$ pseudopotential, commonly referred to as hard-core interaction, to simply simulate electron forces. Implementing both hard-core and attractive interactions in GaAs/AlGaAs heterostructure materials is challenging. Nevertheless, ultra-cold atoms in a rapidly rotating trap present a feasible approach to realizing these attractive interactions~\cite{Cooper,Regnault04,Baranov,Qiu,zxhu}. In this case, the integral form of the interaction energy is expressed as
\begin{eqnarray}\label{eq:Ev1}
    E_{V_1} 
    &=& -{\frac{e^2}{\epsilon l_B} \int dk e^{-k^2} \left[\sum_{m=0}^{N-1}L_{(\beta m)}(k^2/2)\right]^2 k  L_1(k^2/l_B^2)} \nonumber \\
    & & +{\frac{e^2}{\epsilon l_B } \int dk e^{-k^2} \sum_{m=0}^{N-1} \left[L_{\beta m}(k^2/2)\right]^2 k  L_1(k^2/l_B^2) } \nonumber \\
    & & +\frac{2e^2}{\epsilon l_B } \sum_{m = m'+1}^{N-1} \sum_{m'=0}^{N-2} \frac{(\beta m')!2^{\beta m'}}{(\beta m)!2^{\beta m}} \int dk k^{2\beta (m-m')}  \nonumber  \\
    & & e^{-k^2}\left[ L_{\beta m'}^{\beta(m-m')} (k^2/2)  \right]^2 k L_1(k^2/l_B^2),
\end{eqnarray}
in which the three integrals corresponding to direct and two exchange interactions respectively. It should be noted that the $V_1$ interaction comprises equivalent direct and exchange interactions. Moreover, it is well known that the Laguerre function $L_n(x)$ often encounters convergence difficulties when performing integrals, especially over infinite ranges. As the order $n$ increases, the Laguerre polynomials become significantly oscillatory, making the integration more complex over specific intervals. Additionally, as $n$ grows, the zeros of the Laguerre function become closely packed, which might render numerical integration methods less reliable in those regions or require additional techniques to address them. Here, we define $\left[\sum_{m=0}^{N-1}L_{(\beta m)}(k^2/2)\right]^2 k L_1(k^2/l_B^2)=f(k)$ in Eq.~\eqref{eq:Ev1} and determine the coefficient $C_n$ in front of $k^n$, resulting in $f(k)=\sum_{n=0}C_n k^n$. Reformulate Eq.~\eqref{eq:Ev1} as

\begin{align}\label{eq:Ev1_cn}
        E_{V_1} = \frac{e^2}{\epsilon l_B} \sum_{n=0}C_n \Gamma[\frac{1+n}{2}].
\end{align}
This method eliminates the requirement for integrating the Laguerre function, thereby enhancing the accuracy of the calculations. Essentially, the energy for any number of particles $N$ can be obtained using Eq.~\eqref{eq:Ev1_cn}.

\section{Numerical results}
\label{sec:Numerical}

\begin{figure}[H]
    \centering
    \center{\includegraphics[width=8.5cm]  {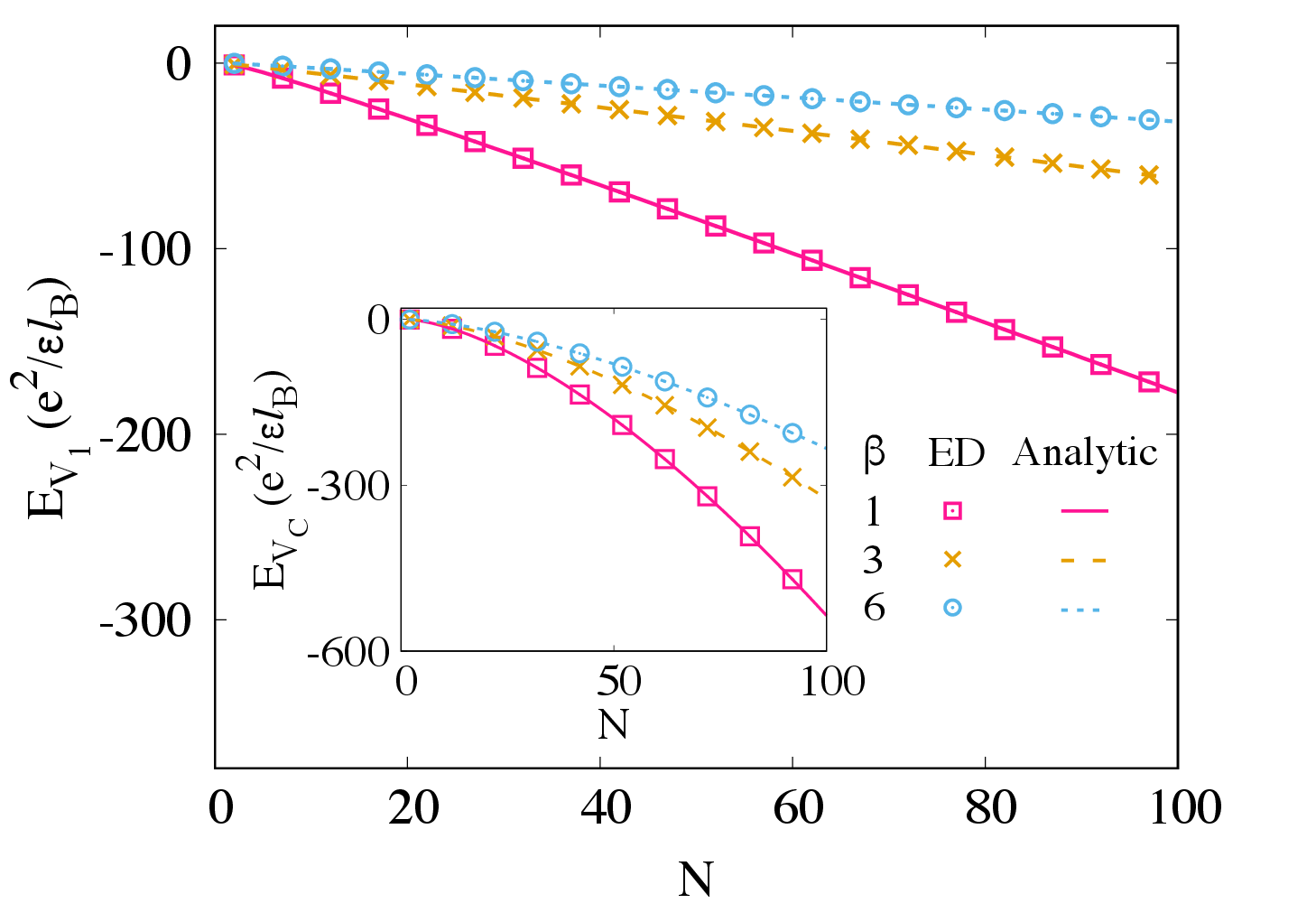}}
    \caption{The dependence between the number of electrons $N$ and $E_{V_1}$ and $E_{V_C}$, respectively. The analytical results are in high agreement with ED. In the same system, both the absolute values of $E_{V_1}$ and $E_{V_C}$ decrease as $\beta$ diminishes, but $E_{V_1}$ decreases at a slower rate.
    }
\label{fig:energy}
\end{figure}

At first, to demonstrate the validity, we compare the analytical integration results with numerical ones from exact diagonalization (ED), applicable to both the Coulomb and $V_1$ model Hamiltonians with various values of $\beta$. We would like to point out that diagonalization is not involved here, as the IQH state possesses a singular basis. In this scenario, we determine the energy directly from the occupation space similarly to how the FQH ground state is derived. As shown in Fig.~\ref{fig:energy}, the numerical findings align closely with the analytical integration. 

The results of the inset in Fig.~\ref{fig:energy} indicate that in the same system, as $\beta$ increases, the absolute value of $E_{V_C}$ decreases, mainly because the increase in $\beta$ leads to an increase in orbital spacing, and the Coulomb interaction is inversely proportional to the relative coordinates of the particles, thus resulting in a decrease in the absolute value of $E_{V_C}$ with increasing $\beta$. The findings in Fig.~\ref{fig:energy} for the ${V_1}$ interaction demonstrate a linear behavior of $E_{V_1}$ with increasing system size in cone geometry. This linear decrease is slower than the Coulomb interaction.

\begin{figure} 
        \centering
        \center{\includegraphics[width=8.5cm]  {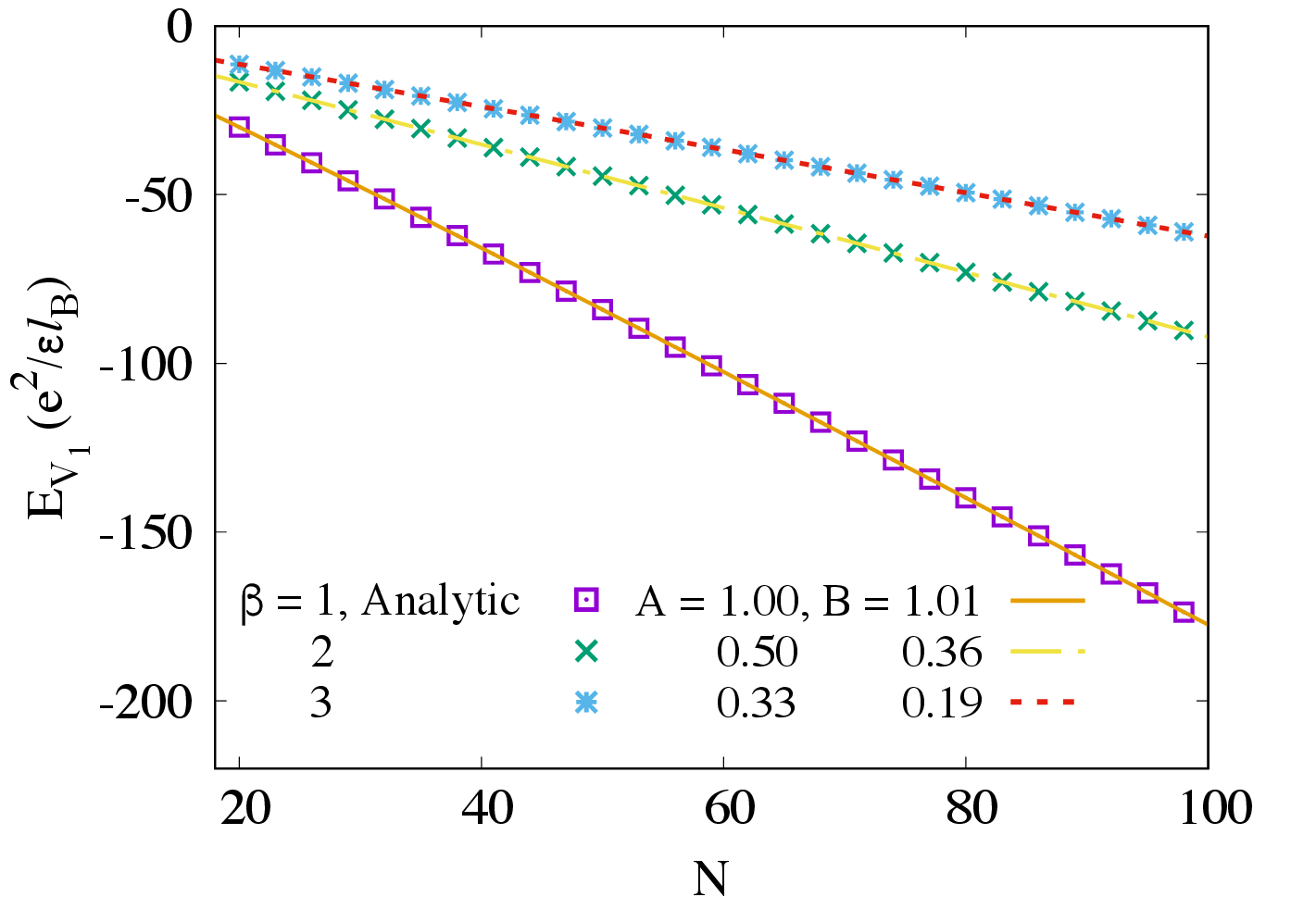}}
        \caption{The variation of $E_{V_1}$ in systems containing 20-100 electrons at $\beta=1-3.$ Use Eq.~\eqref{eq:E_N_cone} to fit the $E_{V_1}$, and obtain $A$ and $B$ under different $\beta$.}
        \label{fig:Ev1fit}
\end{figure}

Similar to disk geometry, in a finite system with $V_1 = -1$ interaction, we divide the energy into bulk energy and interface energy
\begin{align}\label{eq:E_N_cone}
     E_{V_1} = -A2N +B\frac{4}{\sqrt{8\pi^3}}\frac{2\pi}{k_1} +\mathcal{O}(k),
\end{align}
 the first term corresponds to the bulk contribution in the thermodynamic limit, while the second term correspond to the finite-size interface energy contributions. When $\beta=1$, the IQH state for $N$ electrons has radius $\sqrt{2N}$ in units of $l_B$ and is taken as unity. Therefore, $k_1 = 1/\sqrt{2N}$ is the curvature of the interface. Compared to Ref.\onlinecite{Liqi2021}, we have two additional undetermined coefficients, $A$ and $B$, which are associated with $\beta$. Therefore, $A = B = 1$ corresponds to the disk geometry with $\beta = 1$. The results of $E_{V_1}$ can help establish the connection between $A$, $B$, and $\beta$. Taking $\beta = 1-3$ as examples, for a system with $20-100$ electrons, $E_{V_1}$ is shown in Fig.~\ref{fig:Ev1fit}. The coefficients $A$ and $B$ are treated as fitting parameters and fitting is performed at different values of $\beta$ using Eq.~\eqref{eq:E_N_cone}. 

\begin{figure}[H]
        \centering
        \center{\includegraphics[width=8.5cm]  {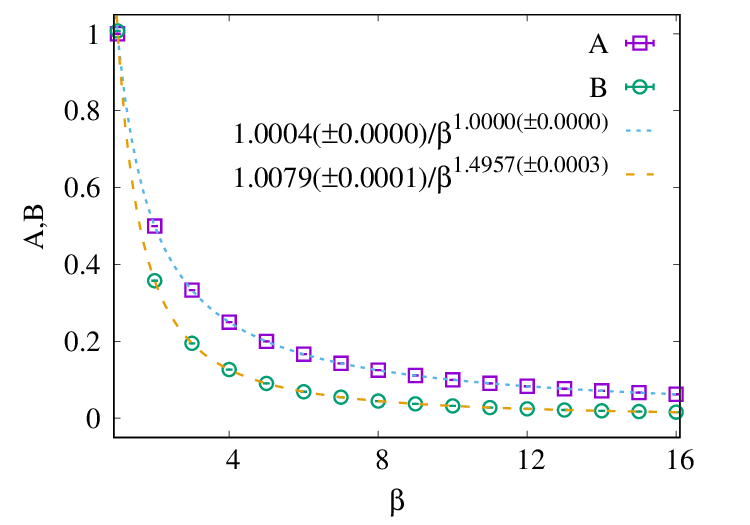}}
        \caption{In systems $N=20-100$, the dependence of the coefficients $A,
        B$ on $\beta$ and mark the corresponding error bars. By performing power fitting for each coefficient, we have derived $A\approx \frac{1}{\beta},B\approx \frac{1}{\beta \sqrt{\beta}}$.}
        \label{fig:fitAB}
\end{figure}

When $\beta = 1$, we get the fitting parameters $A = 1.0$ and $B = 1.01$ which are expected for the disk geometry. If $\beta > 1$, both $A$ and $B$ decrease with an increase in $\beta$. In order to determine the relationship between $A$, $B$, and $\beta$, where $\beta$ ranges from 1 to 16, we illustrate this correlation in Fig.~\ref{fig:fitAB}. Our findings indicate that $A\approx \frac{1}{\beta}$ and $B\approx \frac{1}{\beta \sqrt{\beta}}$. Therefore, according to our fitting results, the interaction energy on a cone could be written as:
\begin{align}\label{eq:V_beta}
   E_{V_1} &= \frac{1}{\beta}(-2N +\frac{1}{\sqrt{\beta}}\frac{4}{\sqrt{8\pi^3}}\frac{2\pi}{k_1})+ \mathcal{O}(k).
\end{align}
The overall factor $1/\beta$ can be interpreted as follows. As cone geometry consists of sectors with an angle of $2\pi/\beta$, representing $1/\beta$ of a full circle, the energy is consequently reduced by a factor of $\beta$. Apart from that, because of string-like characteristics, the interface energy is linked to the entire length of the interface $2\pi\sqrt{2N/\beta}$. Consequently, in contrast to the disk geometry, the interface energy in the cone geometry includes an extra correction factor of $1/\sqrt{\beta}$.

We consider an IQH liquid with attractive interactions which does not require external confinement potentials, where the dynamics of the interface is primarily governed by string tension. According to Ref.~\onlinecite{yangku}, besides the usual bulk Chern-Simons term, the effective action contains the interface contribution that is proportional to the sting tension. In higher order correction (up to 4-order) with consideration of larger deformation, the string tension per unit length $u$ as a function of curvature $k(\beta)$ is $u(k)=u_0+u_2k^2+u_4k^4+\mathcal{O}(k^6)$, where $u_0$ is the coefficient in front of $k^0$, while $u_2$ and $u_4$ are the coefficients of higher-order terms, and the odd terms of $k$ are zero due to the particle-hole symmetry of the two-body Hamiltonian. Taking into account the higher-order terms of curvature, the interface energy, which is the rest energy after substrating the bulk part, is expressed as:
\begin{align}\label{eq:IE}
 E_{I} = (u_0+u_2\frac{\beta}{2N}+u_4\frac{\beta^2}{4N^2})2\pi \sqrt{\frac{2N}{\beta}},
\end{align} 
based on our previous analysis, there exists an overall factor of $1/\beta$ due to the geometry. Therefore, we predict the following relationships for the string tension coefficient: $u_0\propto 1/\beta,u_2\propto 1/\beta^2,u_4\propto 1/\beta^3$. In compliance with Eq.\eqref{eq:IE}, we conducted a fitting process for system sizes $N$ between 50 and 100, which allowed us to determine the values for the parameters $u_0, u_2, u_4$. 

\begin{figure}[H]
        \centering
        \center{\includegraphics[width=8.5cm]  {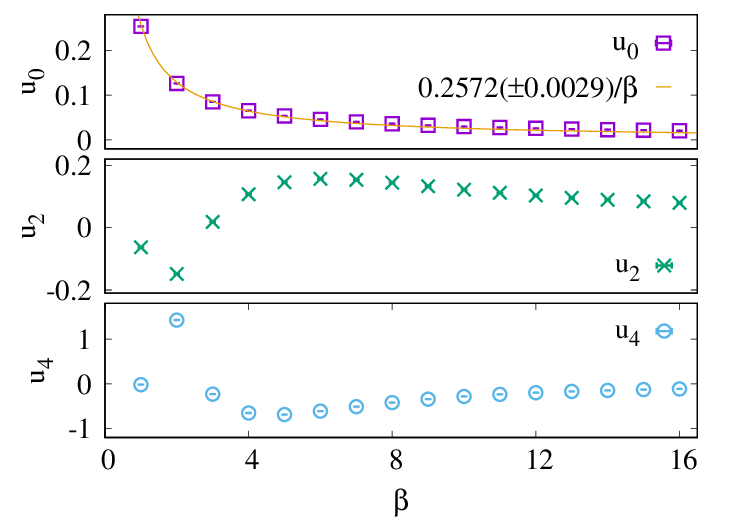}}
        \caption{Under system sizes $N=50-100$, the dependence of the string tension coefficient on $\beta$ is fitted using Eq.~\eqref{eq:IE}, with error bars indicated. When $\beta=1$, we have $u_0=0.253, u_2=-0.063$. $u_0$ has an inverse relationship with $\beta$, while $u_2$ and $u_4$ undergo a notable abrupt change at $\beta=2$.}
        \label{fig:u0_2_4}
\end{figure}

Fig.~\ref{fig:u0_2_4} illustrates the dependence of the string tension coefficients $u_0, u_2, u_4$ on $\beta$, with error bars for the fits indicated in the figure. When $\beta=1$, we find $u_0=0.253$ and $u_2=-0.063$, consistent with the theoretical values derived for the disk geometry up to the quadratic term in Ref.~\onlinecite{yangku}, where $u_0=4/\sqrt{8\pi^3}$ and $u_2=-1/\sqrt{8\pi^3}$. Following this, we carry out a fitting procedure for the parameters $u_0$ and $\beta$ and obtian $u_0=0.2572(\pm 0.0029)/\beta$. This indicates that $u_0\propto 1/\beta$, which aligns with our previous analysis. However, there exist certain deviations between the values of $u_2$ and $u_4$ and our expectations. 

We noticed that at $\beta=2$, the coefficients $u_2$ and $u_4$ of the higher-order terms undergo a sudden change, which reminds us of the contribution of the cone tip angle $(\theta=2\pi/\beta)$ to the entanglement entropy in the cone geometry, as $\theta$ gradually increases from $0$ to $\pi$ (which corresponds to $\beta=2$), the entanglement entropy transitions from a finite value towards zero~\cite{yedan}. In addition, an extremum emerges near $\beta=6$, and previous studies have confirmed that there is a local charge accumulation near the cone tip, and the region close to the cone tip starts to form a CDW phase at this value~\cite{Liqi2024}. These two signals suggest that, in addition to the bulk energy and the interface energy, the contribution of the cone tip should not be overlooked. Fig.~\ref{fig:cone} illustrates that when $\beta > 1$, a concentration of density begins to emerge at the tip of the cone, deviating from the uniformity observed within the bulk density. This may lead to tip energy contributions that are distinct and separate from those of the bulk and interface energies.  Based on this, we will proceed to calculate the cone tip energy and make corrections to the results of $u(k)$.

\begin{figure}[H]
        \center{\includegraphics[width=8.5cm]  {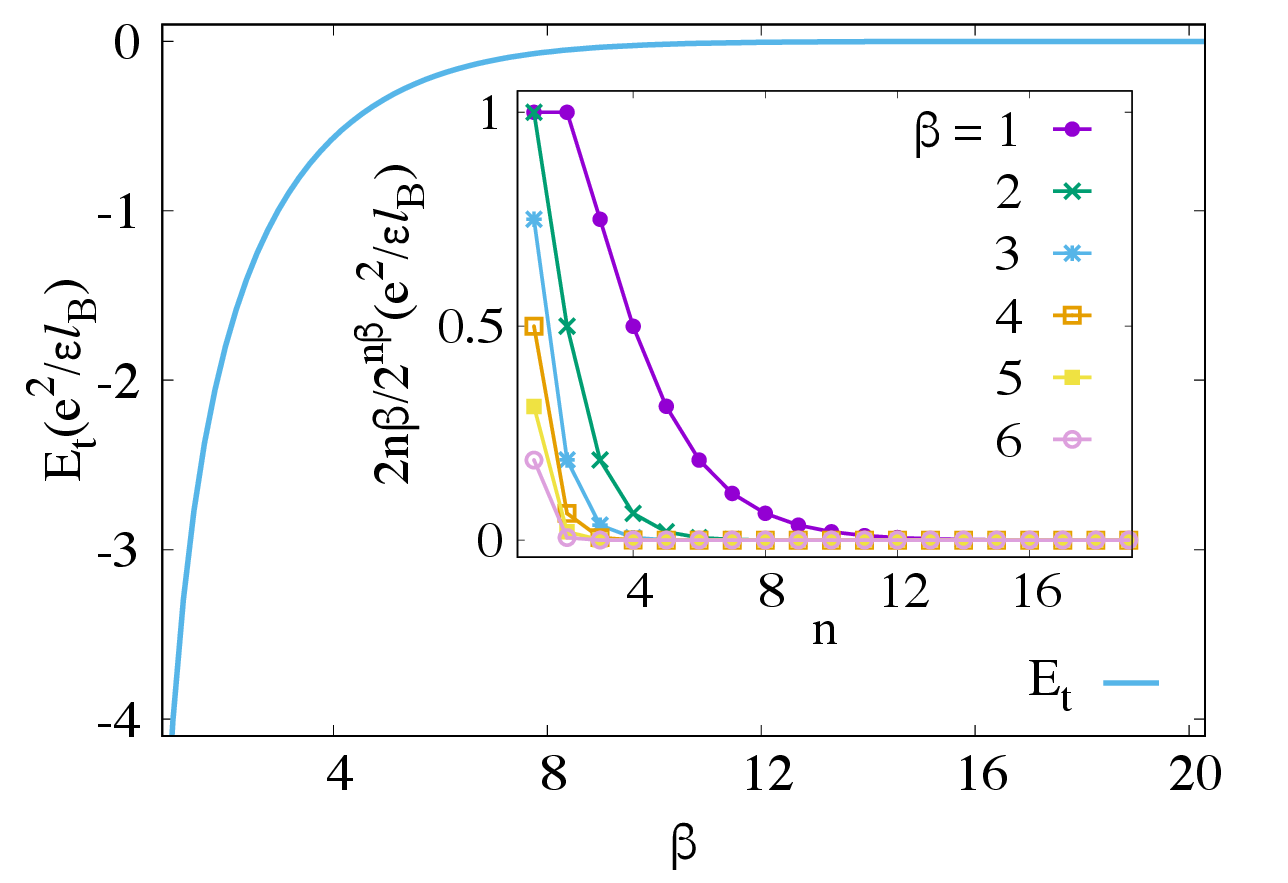}}
        \caption{The tip energy is presented as a function of $\beta$. The magnitude of $E_t$ decreases as $\beta$ increases, which is analogous to the behavior of entanglement entropy near the cone tip. The inset depicts the magnitude of the interaction energy between the 0th orbital and other orbitals.}
        \label{fig:Et}
\end{figure}

We calculated the tip energy $E_t$ using the Hartree-Fock approximation
\begin{equation}\label{eq:Et}
        E_t=\sum_{n}{V_{nmnm}-V_{nmmn}}, 
\end{equation}
in which the matrix element $V_{\{m_i\}}$ is defined as
\begin{eqnarray}\label{eq:V1234}
        V_{\{m_i\}} =& & \frac{e^2}{\epsilon}\int d^2r_1 \int d^2r_2 \psi^*_{\beta m_1}(\vec{r_1})\psi^*_{\beta m_2} (\vec{r_2}) \nonumber \\
        & & V(r_1,r_2) \psi_{\beta m_3}(\vec{r_1})\psi_{\beta m_4}(\vec{r_2}).
\end{eqnarray}
To simplify our analysis, we consider the situation where the tip holds only a single electron in the zeroth orbit, thereby setting $m = 0$ in Eq.~\eqref{eq:Et}. Then the tip energy could be further simplified as:
\begin{equation}\label{eq:Et_s}
        E_t=-\frac{e^2}{\epsilon}\sum_{n=1}^{N-1}\frac{2n\beta}{2^{n\beta}}.
\end{equation}
In the thermodynamic limit with $N \to \infty$, this can be written as $\lim_{{N \to \infty}} E_t = -\frac{e^2}{\epsilon}\frac{2\beta 2^{\beta}}{(2^{\beta}-1)^2}$. We plotted the results of $E_t$ in Fig.~\ref{fig:Et}, as $\beta$ increases, the magnitude of $E_t$ gradually decreases which is quit similar to the behavior of entanglement entropy near the cone tip~\cite{yedan}. Furthermore, the inset graph illustrates that as the orbital index $n$ increases, the interaction strength between the 0th orbital and other orbitals gradually weakens. Additionally, for the same orbital index $n$, the interaction strength decreases as $\beta$ increases. This result is easily understandable, in the cone geometric configuration, each Landau orbital occupies a constant area, leading to a non-uniform rate of spacing variation.  
As the cone angle $(2\pi/\beta)$ decreases, the radial size of the cone tends to increase, accompanied by an increase in the spacing between adjacent Landau orbitals and a decrease in the overlap region between them. We want to point out that the energy contribution at the tip of the cone, denoted $E_t$, is relatively small compared to the total energy of the system, $E_{V_1}$. A meaningful impact from this contribution becomes apparent only when examining the higher-order terms of the interface energy, with the main effects observable in $u_2$ and $u_4$.

\begin{figure}[H]
        \centering
        \center{\includegraphics[width=8.5cm]  {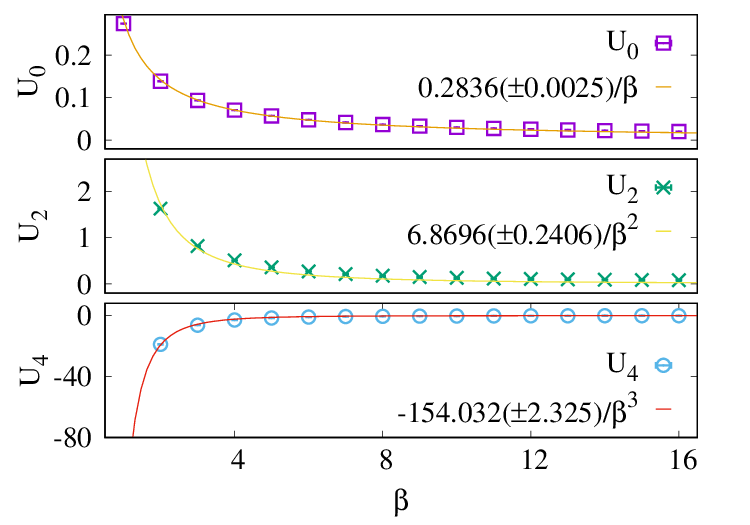}}
        \caption{The dependence of the string tension coefficient $U_0, U_2, U_4$ on $\beta$, obtained through fitting using Eq.~\eqref{eq:Uk}, is represented by different points, with error bars indicated. The fitted curves exhibit the expected behavior satisfying $U_0\propto 1/\beta,U_2\propto 1/\beta^2,U_4\propto 1/\beta^3$.
        }
        \label{fig:U0_2_4}
\end{figure}

At this point, the system's total energy is categorized into three components: bulk, interface, and tip. After considering $E_t$, we have made the necessary corrections to Eq.~\eqref{eq:IE} and redefined the corrected curvature function as $U(k)=U_0+U_2k^2+U_4k^4+\mathcal{O}(k^6)$:
\begin{align}\label{eq:Uk}
       E_{I} = (U_0+\frac{U_2\beta}{2N}+\frac{U_4\beta^2}{4N^2})2\pi\sqrt{\frac{2N}{\beta}}.
\end{align}
After subtracting the tip energy, we still predict the following relationships for the string tension coefficient: $U_0\propto 1/\beta,U_2\propto 1/\beta^2,U_4\propto 1/\beta^3$. Using Eq.~\eqref{eq:Uk}, we fitted the coefficients of $U(k)$ within the range of system sizes $N$ from 50 to 100, and the results are presented in Fig.~\ref{fig:U0_2_4}. We have observed that there is no significant change in the magnitude and trend of $U_0$ relative to $u_0$. For this situation, both $U_2$ and $U_4$ vary monotonically with increasing $\beta$ and do not exhibit a sudden transition.

We conduct a power-law fit for the string tension coefficients $U_0$, $U_2$, and $U_4$. The fitting results show that the relationship of each coefficient with $\beta$ can be expressed as $U_0=0.2836(\pm 0.0025)/\beta, U_2=6.8696(\pm 0.2406)/\beta^2$ and $U_4=-154.032(\pm 2.325)/\beta^3$. Our results indicate that, after considering the contribution of the cone tip energy, the expected behaviors of $U_0\propto 1/\beta,U_2\propto 1/\beta^2,U_4\propto 1/\beta^3$ are satisfied which are in accordance with the predictions made by Eq.~\eqref{eq:Uk}.

\section{Conclusion}
\label{sec:Conclusion}
In this study, we derived analytical formulas for the energy of the electron interaction in IQH systems configured in cone geometry. We formulate the interaction energy as a 1D integral for both the Coulomb and $V_1$ interactions. When the interaction is attractive, the boundary functions as a QH interface. Unlike disk geometry, which only has nonuniform density at the boundary, the cone's apex features a non-uniform density, significantly impacting the total energy. Using the Hartree-Fock approximation, we calculated the contribution from this apex. As $\beta$ increases, the energy of an electron at the tip interacting with others decreases rapidly, mirroring the behavior of the entanglement entropy of the geometric contribution at the tip. Upon considering the tip energy, we found that the energy per unit length of the interface is influenced by its curvature, described by $(C_0+C_2k^2+C_4k^4+\mathcal{O}(k^6))/\beta$ up to a fourth-order expansion where $C_i$ are constants. This result closely corresponds to the previous analysis of effective low energy field theory. Therefore, we conclude that the cone geometry introduces a novel geometry that facilitates the examination of the geometric effects of the quantum Hall interface, offering ease of adjustment.

\acknowledgments
 This work at Chongqing University was supported by the National Natural Science Foundation of China Grant Nos. 12347101 and 12474140, and the Fundamental Research Funds for the Central Universities Grant No. 2024CDJXY022. Q.L. was supported by National Natural Science Foundation of China Grant No. 61988102, the Key Research and Development Program of Guang-dong Province Grant No. 2019B090917007, the Science and Technology Planning Project of Guangdong Province Grant No. 2019B090909011 and Guangdong Provincial Key Laboratory Grant No. 2019B121203002.

\bibliography{biblio_cone.bib}

\end{document}